\documentstyle[12pt,aasms,tighten]{article}

\def\etal{\hbox{et al.\ }$\,$}
\def\simlt{\lower.5ex\hbox{$\; \buildrel < \over \sim \;$}}
\def\simgt{\lower.5ex\hbox{$\; \buildrel > \over \sim \;$}}

\renewcommand{\sun}{\mbox{$_\odot$}}

\received{21 November 1997}
\accepted{12 January 1998}
\slugcomment{\bf To appear on the Astrophysical Journal Letters}

\begin{document}

\title{Measuring the Motion of the Black Hole in GRO J1655$-$40} 

\author{
Roberto Soria\altaffilmark{1,2},
Dayal T.~Wickramasinghe\altaffilmark{2},
Richard W. Hunstead\altaffilmark{3},
Kinwah Wu\altaffilmark{3}}

\altaffiltext{1}{Mount Stromlo and Siding Spring Observatories, 
Institute of Advanced Studies,
Australian National University, Private Bag, Weston Creek Post 
Office, ACT 2611, Australia; roberto@mso.anu.edu.au}
\altaffiltext{2}{The
Australian National University Astrophysical Theory Centre, Canberra, 
ACT 0200, Australia}
\altaffiltext{3}{School of Physics, University of Sydney, NSW 2006, Australia}

\begin{abstract}
\rightskip=\leftskip

We report optical spectroscopic observations of
the Low Mass X-ray Binary GRO J1655$-$40 during 
outbursts in August$-$September 1994 and in June 1996. 
The spectra show emission and absorption lines from the accretion
 disk surrounding the X-ray source, which have enabled us to measure 
the motion of the black hole candidate in this system.
It is the first time among black-hole candidates 
that an emission-line radial velocity curve is seen in 
phase with the expected motion of the primary.
 The projected radial velocity semi-amplitude 
determined for the primary ($K_{X}= 76.2 
\pm 7.5$ km/s),
 combined with estimates of the projected radial velocity semi-amplitude 
of the companion
 star from previous observations 
($K_{C}= 228.2 \pm 3.0$ km/s), 
yields a 95\% confidence limit, derived purely from kinematics, of 
$M_{X} > 5.1 M_{\sun}$ for the mass of the 
compact object in GRO J1655$-$40, confirming its black hole nature.

\end{abstract}

\keywords{accretion, accretion disks -- binaries: spectroscopic 
-- black hole physics -- stars: individual (GRO J1655$-$40)}


\section{INTRODUCTION}

The soft X-ray transient Low-Mass X-ray Binaries have 
proven to be an ideal hunting ground
for stellar-mass black hole candidates (Tanaka \& Lewin 1995).
Among the X-ray binaries commonly discussed as being most likely to harbour 
black holes, all but three (Cyg X-1, LMC X-1, LMC X-3)
belong to this class. Of these, the recently 
discovered GRO J1655$-$40 is particularly 
interesting as well for being a source of superluminal radio 
jets (Zhang \etal 1994; Harmon \etal 1995). Given its optical brightness 
and some evidence of eclipse-like features, it appears to be one of 
the few systems likely to yield a reliable estimate for the mass 
of the collapsed star.

In all but the faintest systems, it is possible to determine 
the projected radial velocity of the companion star $v_{C}$, which 
for a circular Keplerian orbit is given by:
\begin{displaymath}
v_{C}(t) = K_{C}\, \cos \left[2\pi\phi (t)\right] + \gamma,
\end{displaymath}
where 
$\gamma$ is the systemic velocity and $\phi (t)$ is the orbital phase.
The measurement of the orbital period $P$ and of the projected radial 
velocity semi-amplitude $K_{C}$ leads to an estimate of 
the optical mass function:
\begin{displaymath}
f_{C} \equiv \frac{(M_{X}\sin i)^{3}}{(M_{C} + M_{X})^2} \equiv 
\frac{M_{X} \sin^3 (i)}{(1+q)^2} = \frac{PK_{C}^3}
{2\pi G}  
\end{displaymath}
(van Paradijs \& McClintock 1995; 
Shapiro \& Teukolsky 1983), where $q \equiv M_{C}/M_{X}$, 
$M_{X}$ and $M_{C}$ are the masses of the compact object and of the 
companion star respectively, and $i$ is the
inclination angle of the orbital plane.

An estimate of $M_{C}$ can usually be obtained from the spectral type 
and the luminosity of the companion.
Modelling the ellipsoidal intensity variations of the companion 
during a low state is another way to determine $q$ and $i$.
These methods have been applied with considerable success 
to GRO J1655$-$40 (Orosz \& Bailyn 1997), yielding $K_{C}= 228.2 \pm 3.0$ km/s,
$\gamma = -142.4 \pm 1.6$ km/s,
$f_{C} = (3.24 \pm 0.09)$ $M_{\sun}$, $M_{C} = (2.34 \pm 0.12)$ $M_{\sun}$, 
$i = (69.5 \pm 0.08)\deg$ \, and $M_{X} = (7.02 \pm 0.22)$ $M_{\sun}$.

Nevertheless, given the uncertainties in estimating the mass of the 
companion star 
from its spectral type, and in modelling the ellipsoidal variations in 
the presence of an accretion disk, an independent mass estimate is 
desirable.
This can be achieved by determining the radial velocity curve of 
the X-ray source. In the case of NS X-ray binaries, one 
can often measure the Doppler shift of coherent X-ray 
pulses from the central object, whereas in BH binaries one has to resort to 
optical studies of the accretion disk (Orosz \etal 1994).
Measuring the radial velocity curve of the primary would allow 
us to determine the X-ray mass function:
\begin{displaymath}
f_{X} \equiv \frac{(M_{C}\sin i)^{3}}{(M_{C} + M_{X})^2} \equiv 
\frac{M_{C} \sin^3 (i)}{(1+1/q)^2} = \frac{PK_{X}^3}
{2\pi G}.  
\end{displaymath}
All the physical parameters would then be determined as a function of $i$. 
For example, $q \equiv M_{C}/M_{X} = K_{X}/K_{C} =  
\left(f_{X}/f_{C}\right)^{1/3}$, 
and $M_{X} = f_{C} \, (1+q)^2 \, (\sin i)^{-3}$.

In the following sections we discuss the results of 
spectroscopic observations carried out during outbursts in 1994 and 1996. 
In particular we focus on the radial velocity variations of the
He{\small{ II}} $\lambda 4686$ emission line, from which we infer the velocity 
curve of the primary. We also examine the fluorescence lines 
N{\small{ III}} $\lambda \lambda 4641, 4642$, which appear to come from 
a hot spot near the outer edge of the accretion disk.

\section{OBSERVATIONS AND RESULTS}

GRO J1655$-$40 was observed briefly, as a target of 
opportunity, on each night from August 30 to 
September 4, 1994, with the RGO spectrograph and Tektronix 1K $\times$ 1K 
thinned CCD on the 3.9m Anglo-Australian Telescope. Spectra of 
two wavebands were obtained: the $6278-6825$ \AA\ band 
(centred on H$\alpha$), 
and the $4432-5051$ \AA\ band (covering N{\small{ III}}, He{\small{ II}} 
and H$\beta$); 1200 g/mm gratings were used for both bands, oriented 
with the blaze direction towards the 25 cm camera, and the resolution 
was 1.3 \AA\  FWHM. 
More extensive observations were carried out on June 8$-$12, 1996, using 
the Double Beam Spectrograph on the 2.3m ANU 
telescope at Siding Spring Observatory, with 1200 g/mm gratings 
for both the blue ($4150-4990$ \AA) and the red ($6300-7200$ \AA) 
spectral regions (resolution 1.2 \AA \ FWHM); the detectors used were 
SITE 1752 $\times$ 532 CCDs in both arms of the spectrograph. 
Conditions were not photometric 
in June 1996, and we normalized the spectra to the continuum level, 
using the ``continuum'' task in IRAF.
Both in August-September 1994 and in June 1996 the optical spectrum 
was dominated by a bright accretion disk.
In both sets of data we observed strong, double-peaked H$\alpha$ 
and He{\small{ II}} $\lambda 4686$ emission lines, and much weaker emission 
from H$\beta$, H$\gamma$ and He{\small{ I}} $\lambda 6678$.
Strong emission was consistently seen
from the Bowen fluorescence lines of N{\small{ III}} $\lambda 
\lambda 4634, 4641, 4642$, with weak emission lines of C{\small{ III}}, 
N{\small{ II}}, O{\small{ II}} identified at some orbital phases. 
The H{\small{ I}} Balmer lines also show a broad absorption trough, 
much broader than the emission line (Hunstead, Wu, \& Campbell-Wilson 1997). 
Some narrow absorption lines from the spectrum of the companion star 
are also detectable, although they tend 
to be swamped by the bright continuum emission from the disk. 
Portion of the He{\small{ II}} region obtained 
by coadding the data taken on June 8, 1996 is shown in Figure 1; 
portion of the H$\alpha$ spectrum obtained by coadding all June 1996 data 
is shown in Figure 2.

\begin{figure}
\epsfxsize=170mm\epsfbox{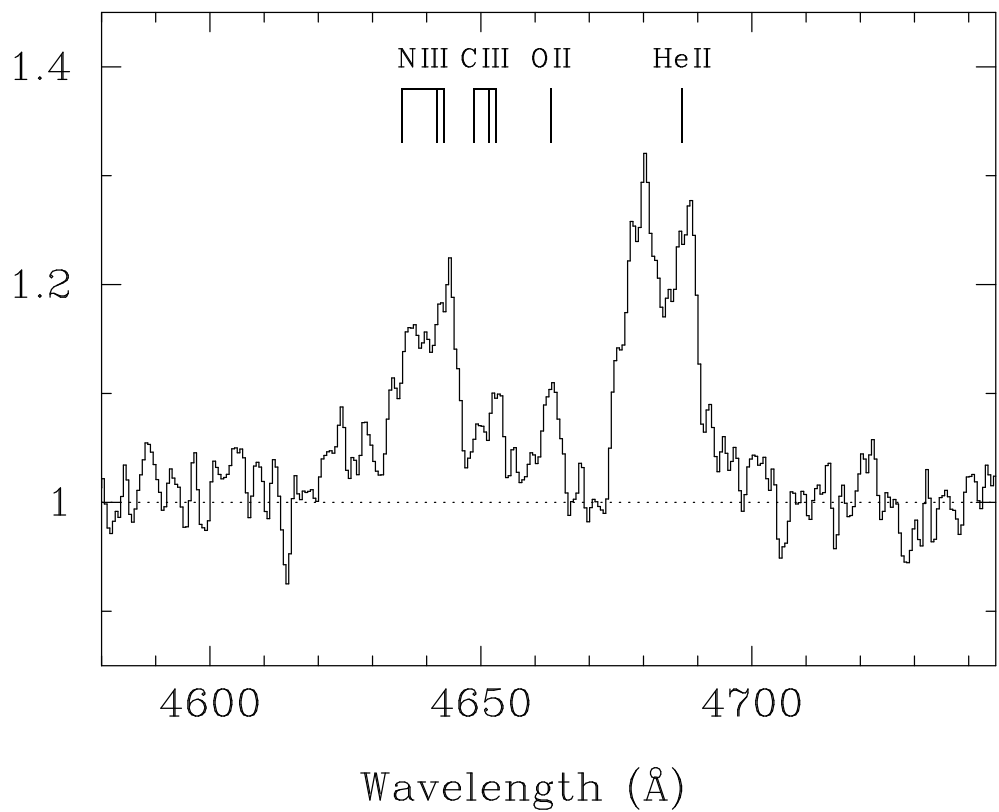} 
\caption{Portion of the blue spectrum of GRO J1655$-$40 obtained 
by coadding the 
spectra taken on June 8, 1996 (orbital phases $0.07 < \phi < 0.17$) 
with the ANU 2.3m telescope at Siding Spring Observatory. 
Wavelengths are vacuum heliocentric and the intensity is 
normalized to the continuum, shown as a dotted line.
Lines clearly identifiable are: 
He{\small{ II}} $\lambda 4686$; 
N{\small{ III}} $\lambda \lambda 4634, 4641, 4642$; 
C{\small{ III}} $\lambda \lambda 4647, 4650, 4651$;
O{\small{ II}} $\lambda 4662$.}
\end{figure}

\begin{figure}
\epsfxsize=170mm\epsfbox{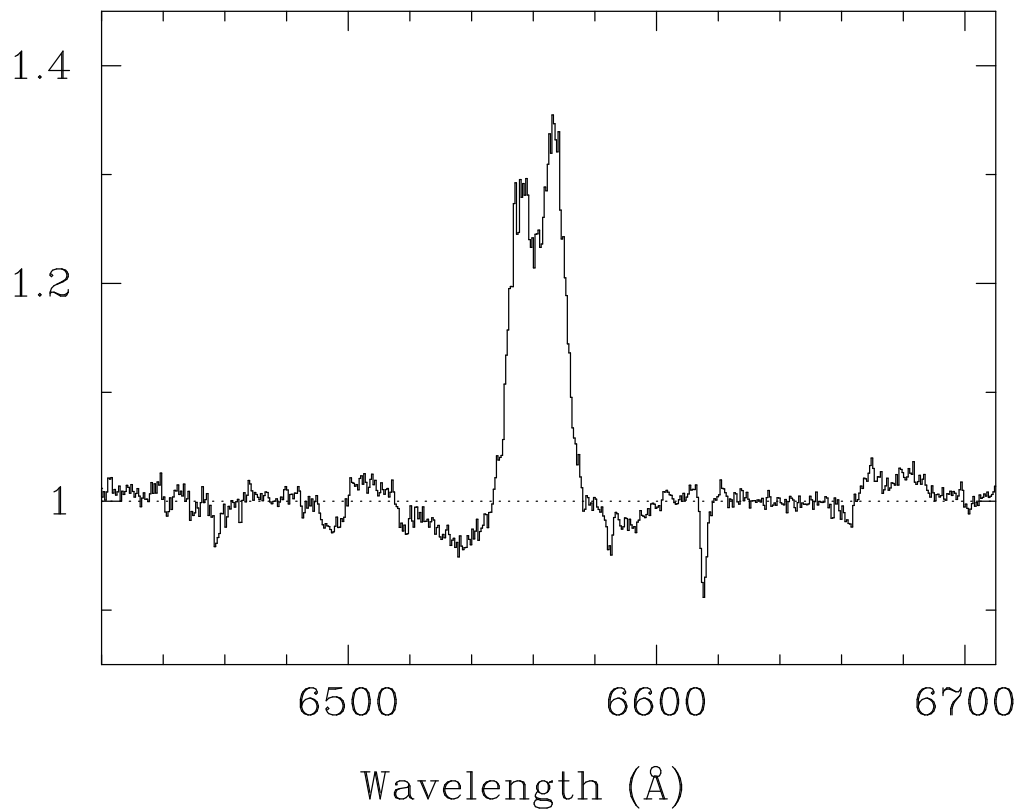} 
\caption{Portion of the red spectrum of GRO J1655$-$40 obtained by coadding 
all the spectra taken in June 1996
with the ANU 2.3m telescope at Siding Spring Observatory; as in Fig.\ 1,
wavelengths are vacuum heliocentric and the intensity is 
normalized to the continuum.
The double-peaked H$\alpha$ $\lambda 6562$ emission line is prominent;
the peak-to-peak separation is $\sim 500$ km/s.
A much weaker, double-peaked He{\small{ I}} $\lambda 6678$ line 
can also be identified, and the blend Fe{\small{ I}} 
$\lambda \lambda 6495, 6496, 
6499, 6502$ is visible in absorption. The absorption line at
6613 \AA\ is interstellar.}
\end{figure}

\section{THE RADIAL VELOCITY CURVE OF THE PRIMARY}

Radial velocity variations of disk emission lines  
were measured by Haswell \& Shafter (1990) for the X-ray binary 
A0620$-$00, and by Orosz \etal (1994) for A0620$-$00 and GRS 1124$-$68,
in each case to determine the mass of the primary. 
One method for determining the central position of the line is to 
convolve the data with two identical Gaussian bandpasses. When the 
counts in each bandpass are equal, the midpoint between the two Gaussians 
is the wavelength of the spectral line. This method assumes that the 
observed line profile is symmetric. 
Another way of finding the line center is to fit the theoretical profile 
expected for a line emitted from the surface of a geometrically thin disk 
(Smak 1981; Horne \& Marsh 1986), either to both wings and both peaks, or 
only to the red wing and peak.
By fitting sine curves to the H$\alpha$ radial velocity curves for 
A0620$-$00 and GRS 1124$-$68, 
Orosz \etal (1994) found that the semi-amplitudes are consistent with 
the mass ratios previously estimated for the two systems, but that there is 
in both cases a phase shift of $\sim 40\deg$ from the expected radial 
velocity of the primary.

In our study of GRO J1655$-$40 we focus 
on the velocity variations of 
He{\small{ II}} $\lambda 4686$, for which we have the best phase coverage.
The velocity variations were determined from the midpoint 
of the steep line wings at 1/4 of the maximum intensity above the continuum, 
not by fitting to the peaks. The emission in the peaks comes from 
the outer rings of the disk, where emission from a hot spot and 
tidal effects causing deviations from keplerian rotation can be 
significant; the wings are emitted at smaller 
radii and are probably a more reliable indicator of the line position. 
Errors were estimated 
by measuring the midpoint near the continuum level and at 
half maximum. These three values do not differ by more than $0.5$ \AA \ 
(32 km/s) in any spectrum.

The H$\alpha$ emission line has a higher signal-to-noise ratio than 
He{\small{ II}} $\lambda 4686$, but it cannot be used for an estimate 
of the orbital velocity of the disk, because the underlying broad, often 
asymmetric absorption does not allow an accurate determination 
of a mid-point position from the line wings.
Furthermore, unlike He{\small{ II}} $\lambda 4686$, the H$\alpha$ emission 
line profile is strongly asymmetric: the surface emissivity of the disk 
appears to be a function of the azimuthal coordinate, and the 
absorption line from the companion star might also be present. For these 
reasons, the velocity curve of the H$\alpha$ line does not reflect 
the orbital motion of the disk. We will discuss the H$\alpha$ line profile 
in a paper currently in preparation.

\begin{figure}
\epsfxsize=145mm\epsfbox{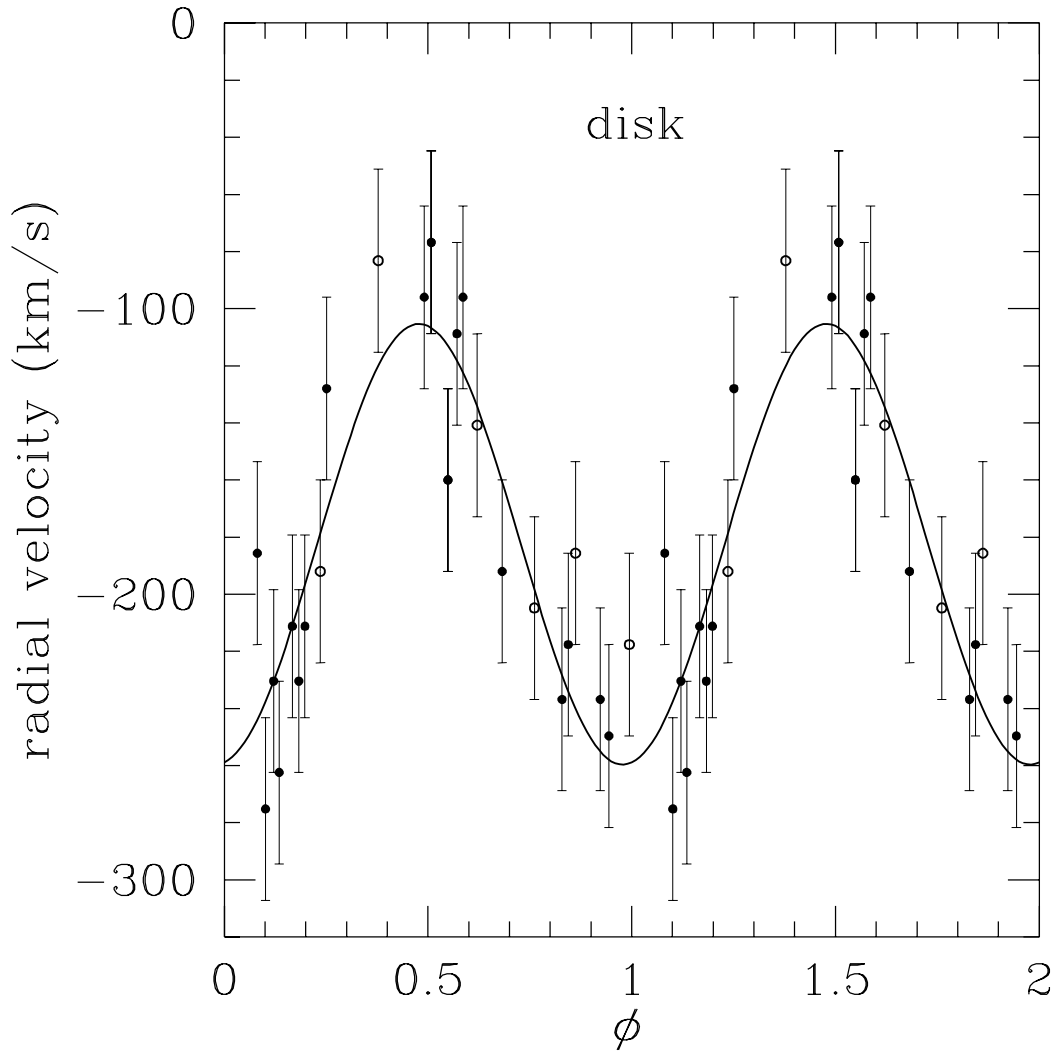}  
\caption{Radial velocity curve for the He{\small{ II}} $\lambda 4686$ 
emission line from the June 1996 (filled circles), and
August-September 1994 data (open circles).
Following Orosz \& Bailyn (1997), we have defined $\phi = 0$ at 
the time of maximum 
redshift of the companion star according to their ephemeris. 
The best fit for the 
radial velocity curve of the disk lines (calculated only from the 1996 data) 
is in antiphase with respect to the star and has a semi-amplitude
$K_{X} = 76.2 \pm 7.5$ km/s.}

\end{figure}

Figure 3 shows the best sinusoidal fit to the 
measured heliocentric radial velocity in the 1996 dataset (filled circles). 
We obtain: 
\begin{displaymath}
v_{X}(t) = 76.2 \cos \left\{2\pi\left[\phi (t) + 0.524\right]\right\} 
- 182.5 \qquad \mbox{km/s}.
\end{displaymath}
The six positions measured in 1994 for the same line 
have also been plotted (open circles) but were 
not included in the calculation of the best fit, because the line profiles 
and equivalent widths appear very different from the 1996 spectra, and 
much more variable from phase to phase, as we shall discuss in more 
detail elsewhere. The radio and X-ray 
behaviour in 1996 was also very different from 1994.
The orbital period has been assumed to be $P = 2.62157$ days, 
and the phase $\phi = 0$ corresponds to the time of maximum redshift 
of the companion star, as determined 
by Orosz \& Bailyn (1997) when the system was in quiescence.
According to their ephemeris, the time of inferior conjunction of the 
secondary star is HJD $244\:9838.42(09) + 2.621(57) \times \mbox{N}$.
The phase shift of $(189 \pm 20) \deg$ between the two radial velocity 
curves is consistent with their being in antiphase and strongly suggests that 
orbital motion of the disk is responsible for the effect, rather 
than, for example, anisotropic emission.
A different ephemeris has recently been proposed by 
van der Hooft \etal (1997), who determine the 
time of inferior conjunction of the 
secondary star at HJD $244\:9838.4198(52) + 2.62168(14) 
\times \mbox{N}$; this would imply a phase shift 
between the radial velocity curves of the disk and the companion star 
of $(191 \pm 20) \deg$, fully consistent with our interpretation. 
If perfect antiphase is demanded in fitting the velocity curve, we 
obtain 
\begin{displaymath}
v_{X}(t) = 72.1 \cos \left\{2\pi\left[\phi (t) + 0.500\right]\right\} 
- 181.8 \qquad \mbox{km/s},
\end{displaymath}
consistent with the previous result.

We also note that: \\
a) the systemic velocity inferred from the He{\small{ II}} 
radial velocity curve ($-182.5 \pm 5.5$ km/s)
is significantly larger (in absolute value) than that 
determined by Orosz \& Bailyn (1997) from the orbital motion of the companion 
star ($-142.4 \pm 1.6$ km/s). In other words, the 
disk emission lines appear systematically blue-shifted by $40$ km/s; and\\
b) the separation between the two peaks in the He{\small{ II}} emission line 
is always $\simlt 650$ km/s (Figure 1).  The corresponding 
separation in the H$\alpha$ 
emission line is always $\simlt 550$ km/s in the June 1996 spectra 
(Figure 2), and always $\simlt 350$ km/s in the August-September 1994 spectra.
It is generally assumed that in a binary system with mass ratio $q > 0.25$ 
the accretion disk around the primary cannot grow any larger than 
the tidal truncation radius $r_d$ (Paczy\'{n}ski 1977; Whitehurst 1988; 
Osaki, Hirose \& Ichikawa 1993), approximately given by 
$r_d = 0.60 \, a /(1+q)$ for $0.03 < q < 1$ (Warner 1995), where $a$ is 
the separation between the centers of mass of the 
binary components. 
Using the orbital parameters derived in this section 
and in Orosz \& Bailyn (1997), we derive 
$a \, \sin i = (11.0 \pm 0.3) \cdot 10^{11}$ cm, and 
$r_d \, \sin i = (4.9 \pm 0.2) \cdot 10^{11}$ cm, 
corresponding to a minimum value of the projected radial velocity 
of the outer rim of the disk 
$v_{\mbox{{\scriptsize rim}}} = 391 \pm 13$ km/s. 
Therefore the peak-to-peak separation 
in the emission lines is always less than twice the projected radial velocity 
of the rim.

Both phenomena are very common, not only among BH candidates but also in 
NS binaries and in cataclysmic variables, and may be interpreted 
as evidence that He{\small{ II}} $\lambda 4686$ 
is emitted in a wind region just above the disk 
surface (Murray \& Chiang 1997). A similar situation may apply to the much 
more massive BH accretion disks in QSOs, where different 
emission lines are found to be systematically blue-shifted with respect to the 
systemic redshift. The magnitude of the blueshift is roughly in order 
of increasing ionization, with the most extreme shift 
being for He{\small{ II}} (Tytler \& Fan 1992).

We must also consider the possibility of 
tidal deformation of the accretion disk by 
the companion star. It is well known, both theoretically (Paczy\'{n}ski 1977) 
and observationally (Marsh, Horne, \& Shipman 1987), 
that if the outer rings of the disk are tidally deformed, 
the observed amplitude of 
the velocity variations of its lines is slightly larger 
than the real amplitude of motion 
of the primary. Since we are measuring the position of the wings (emitted 
from inner regions of the disk, 
$r \simlt 0.5 \, r_{\mbox{{\scriptsize rim}}}$), we expect this tidal
effect to be small, i.e., the true velocity of the primary may have 
been over-estimated by $\simlt 5$\%.

Neglecting tidal effects, we conclude that 
$K_{X} = 76.2 \pm 7.5$ 
km/s, which implies a mass ratio of $q = 0.334 \pm 0.033$.
The X-ray mass function 
is then $f_{X} = (0.121 \pm 0.036) M_{\sun}$, giving 
\begin{displaymath}
M_{X} = \frac{5.77 \pm 0.33}{(\sin i)^3} \, M_{\sun}.
\end{displaymath}
Even without any assumptions on the inclination angle, 
this result shows that $M_{X} > 5.1 M_{\sun}$ (95\% 
confidence limit), well above the Chandrasekhar limit. 
Our result is fully consistent with the mass ratio derived by 
Orosz \& Bailyn (1997) and van der Hooft \etal (1997).

\section{DETECTION OF THE HOT SPOT?}

In our June 1996 spectra
the Bowen fluorescence emission lines [see McClintock \& Canizares (1975), 
and Schachter, Filippenko, \& Kahn (1989) for a discussion on their 
origin and significance] 
N{\small{ III}} $\lambda 4634$ and the blend N{\small{ III}} $\lambda \lambda 
4641, 4642$ are narrow and single-peaked; their profiles suggest
that the emission region is localized. We measured their positions by 
fitting Gaussian profiles.
The observed radial velocities 
for N{\small{ III}} $\lambda \lambda 
4641, 4642$ are shown in Figure 4: the dashed line is the projected radial 
velocity of the companion star determined by Orosz \& Bailyn (1997). 
It is possible that the Bowen lines are emitted in the irradiated atmosphere 
of the secondary, but their velocity semi-amplitude
appears to be larger than $K_{C}$. If the emitting region is on a 
Keplerian orbit, it must therefore be 
located {\it inside} the star's orbit.

\begin{figure}
\epsfxsize=155mm\epsfbox{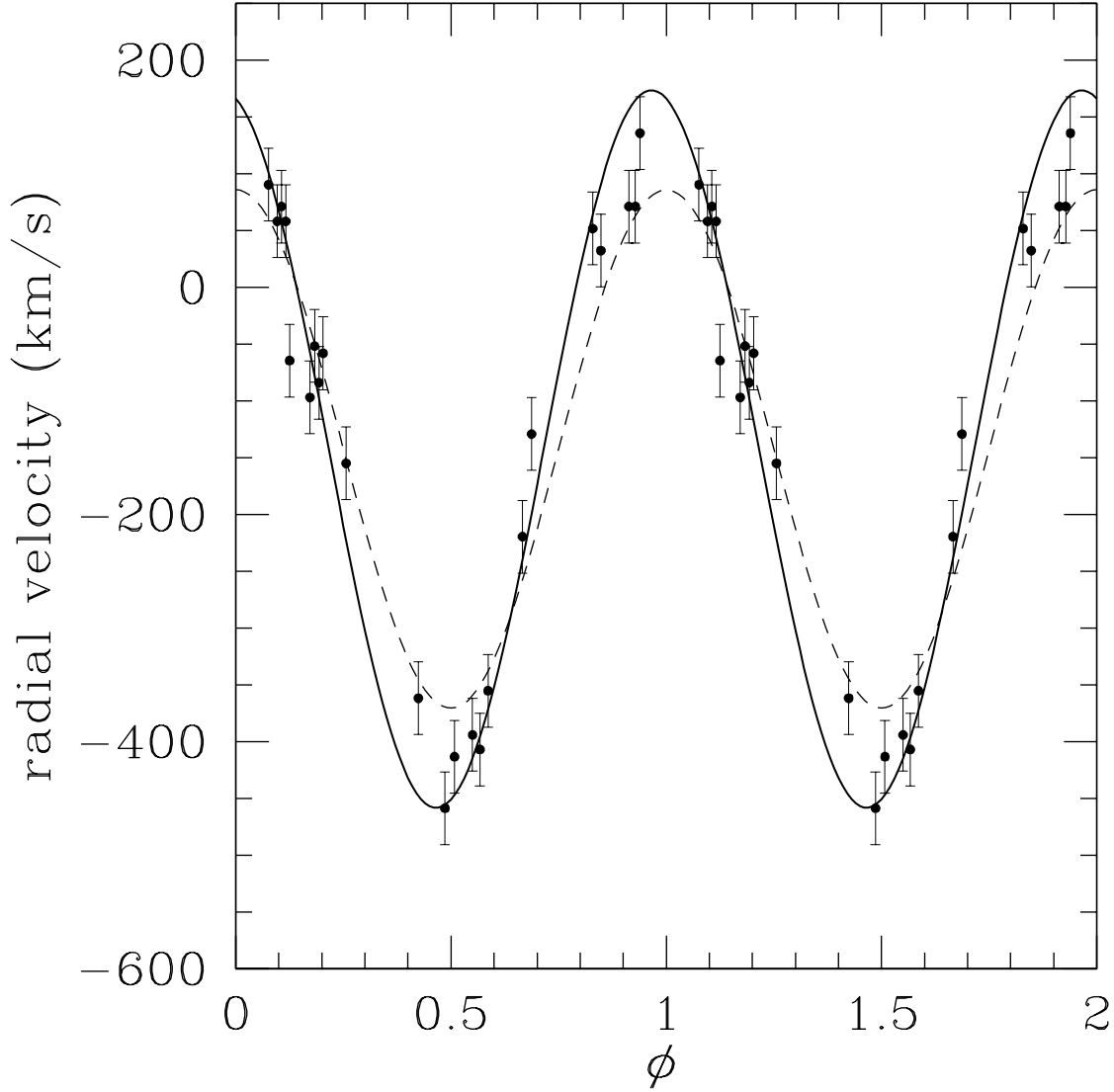} 
\caption{Radial velocities for the narrow Bowen fluorescence lines 
N{\small{ III}} $\lambda \lambda 4641, 4642$ plotted as a function of 
orbital phase. 
The dashed line is the velocity curve of the companion star as determined 
by Orosz and Bailyn.
The solid line is the theoretical projected 
radial velocity of a hot spot on the rim of a thin, circular, 
Keplerian accretion disk truncated at the tidal radius 
$r_{d} \simeq (4.9 \cdot 10^{11} \mbox{ cm})/\sin i$, and 
was calculated by using the mass of the primary and its 
orbital parameters derived from the radial velocity curve of 
He{\small{ II}} $\lambda 4686$ (Fig.\  3).
The hot spot was assumed to be $10\deg$ in front of the 
companion star.}
\end{figure}

An alternative possibility is that the N{\small{ III}} lines originate from 
the hot spot where the accretion stream from the star joins 
the accretion disk, and such an interpretation is indeed consistent 
with the parameters that we have deduced for this system.
The solid line plotted in Figure 4 is not a fit to the data: it is instead 
the theoretical projected radial velocity
of a spot on the rim of the disk, leading the star by $10\deg$.
We have assumed here a circular, thin, Keplerian disk truncated at 
the tidal radius $r_d$ calculated in Section 3, and 
we have made the simplifying assumption that the velocity of the spot will 
be the vector sum of the systemic velocity, the orbital velocity of the 
primary with its accretion disk around the center of mass of the system, 
and the Keplerian velocity of rotation around the primary.
 
This simple model suggests that the Bowen lines originate either from 
the hot spot (in that case, the outer edge of the disk must reach the 
tidal truncation radius) or from the inner part of the accretion stream.

\section{CONCLUSIONS}

We conclude that the broad He{\small{ II}} emission line in GRO J1655$-$40 
shows velocity variations that can be interpreted as being due to the 
orbital motion of the primary (and hence of its accretion disk). 
The observed amplitude of the variations 
allows us to find a 95\% confidence limit of $M_{X} > 5.1 M_{\sun}$ for 
the mass of the collapsed object, thus firmly establishing GRO J1655$-$40 
as a black hole binary system {\it from its kinematics alone}; this is in 
good agreement with the mass $M_{X} = (7.0 \pm 0.2)$ $M_{\sun}$ estimated 
by Orosz \& Bailyn (1997).
We also notice that while some of the emission lines, such as He{\small{ II}} 
and H$\alpha$, are double-peaked and likely to come from the whole disk, 
others, such as the Bowen lines of N{\small{ III}}, seem to come from a 
localized area in the outer region of the disk.

We thank the referee for his careful reading of the paper and helpful 
feedback.
R.~S. also thanks Jerome Orosz for interesting comments and discussions;
R.~W.~H. acknowledges financial assistance from the Australian Research 
Council; K.~W. acknowledges support from the ARC through an Australian 
Research Fellowship.

\end{document}